\documentclass{article}

\usepackage{arxiv}

\usepackage[utf8]{inputenc} 
\usepackage[T1]{fontenc}    
\usepackage{lmodern}        
\usepackage{hyperref}       
\usepackage{url}            
\usepackage{booktabs}       
\usepackage{amsfonts}       
\usepackage{nicefrac}       
\usepackage{microtype}      
\usepackage{graphicx}

\title{Estimating Excess COVID-19 Infections with Nonparametric Self-Exciting
Point Processes}

\author{
    Peter Boyd
   \\
    Department of Statistics \\
    Oregon State University \\
  Corvallis, Oregon \\
  \texttt{\href{mailto:boydpe@oregonstate.edu}{\nolinkurl{boydpe@oregonstate.edu}}} \\
   \And
    James Molyneux
   \\
    Department of Statistics \\
    Oregon State University \\
  Corvallis, Oregon \\
  \texttt{\href{mailto:molyneuj@oregonstate.edu}{\nolinkurl{molyneuj@oregonstate.edu}}} \\
  }

\usepackage{tabularx}
\usepackage{booktabs}
\usepackage{natbib}
\usepackage{amsmath}
\usepackage{cite}

\begin{document}
\maketitle

\def\tightlist{}

\begin{abstract}
The COVID-19 pandemic has led to a vast amount of growth for statistical
models and methods which characterize features of disease outbreaks. One
class of models that came to light in this regard has been the use of
self-exciting point processes, wherein infections occur both ``at
random'' and also more systematically from person-to-person
transmission. Beyond the modeling of the overall COVID-19 outbreak, the
pandemic has also motivated research assessing various policy decisions
and event outcomes. One such area of study, addressed here, relates to
the formulation of methods which measure the impact that large events or
gatherings of people had in the local areas where the events were held.
We formulate an alternative approach to traditional causal inference
methods and then apply our method to assessing the impact that then
President Donald Trump's re-election campaign rallies had on COVID-19
infections in areas where the rallies were hosted. By incorporating
several adaptions to nonparametric self-exciting point process models,
we estimate both the excess number of COVID-19 infections brought on by
the rallies and the duration of time in which these excess infections
persisted.
\end{abstract}


\section{Introduction}

The global COVID-19 pandemic has resulted in an explosion of research
surrounding the statistical modeling of epidemics (see for example
\citep{Malloy2021}, \citep{Bonalumi2020}). The pandemic has also yielded
research related to how events (\citep{Dhaval2021},
\citep{Hadianfar2021}) or government policies (\citep{Wang2021},
\citep{Sang2020}) have impacted various aspects of the outbreak, such as
infection and hospitalization rates, case fatality rates, etc.

\emph{Self-exciting point processes} are one class of models that have
seen increased use in the spatio-temporal modeling and forecasting of
the COVID-19 pandemic \citep{Bertozzi2020, Browning2021}. Initially
developed by \citep{hawkes1971}, this model class has been utilized
extensively in the study of earthquakes (\citep{ogata1988},
\citep{ogata1998}) and has also been applied to areas such as wildfires
\citep{Peng2005}, invasive species \citep{Balderama2012}, mass shootings
in the United States \citep{boyd}, and disease spread (\citep{Park2020},
\citep{Unwin2021}. The use of self-exciting point processes for the
modeling of the COVID-19 virus is a natural fit as the model mimics the
contagion process in a simple, yet realistic manner. A person becomes
infected from a (potentially) unknown source, and they may then spread
the virus to others for some period of time, over which the level of
contagion gradually decays.

In addition to the initiation of the COVID-19 pandemic, the year 2020
also featured a presidential election in the United States in which the
incumbent, President Donald Trump, sought re-election against challenger
Joseph Biden. President Trump held five presidential campaign rallies
during the beginning phases of the COVID-19 outbreak, from February 11
to May 20, before temporarily suspending future rallies. After a brief
hiatus, the rallies continued, beginning on June 20, 2020 in Tulsa,
Oklahoma. These campaign rallies are of interest as they were typically
large in size, with participation ranging from hundreds to over 30,000
attendees \citep{Lock2020}, with little social distancing or
mask-wearing by the participants \citep{npr}.

A central question to address is how these rallies impacted the number
of people infected with the virus in counties where the rallies took
place. Previous researchers have sought to address this question with
the use of causal inference modeling \citep{bernheim2020_real}, whereby
the authors compare the number of infections observed with a
counterfactual model which predicts the number of infections that might
have occurred had the rallies not taken place.

Some COVID-19 studies (not related to campaign rallies) also rely on the
use of the synthetic control method (SCM) to investigate factors
impacting the spread of COVID-19, such as school re-openings
\citep{alfano2021} or local air quality \citep{cole2020}. SCM was
developed to allow for the evaluation of an intervention by comparing
the intervention (treatment) group to a weighted combination of groups
that received no treatment as a proxy for a control group
\citep{abadie2003} and has emerged as an especially pertinent analysis
approach for political science and epidemiology studies
\citep{rehkopf2018}. Implementing SCM can be a challenge, however, as
the method assumes similarity among treatment and control units, a lack
of contamination from the treatment unit into the control units, and a
lack of external ``shocks'' that may impact the control units
\citep{bouttell2018}. Furthermore, the ability to verify these
assumptions requires vast field-specific knowledge of the data, and
there exists no consensus on what constitutes a sufficient amount of
similarity between treatments and controls.

In this article, we investigate the impact of President Trump's rallies
on COVID-19 outbreaks with the use of a nonparametric variant of the
self-exciting point process model. In doing so, we develop statistical
methods which can be used in place of traditional causal inference
methods without the need to specify synthetic controls. In Section
\ref{Methods}, we provide background information on self-exciting point
process models and the adaptations to the model we employ to estimate
the excess number of cases induced by an event and the duration of the
excess. We also describe how the model is estimated and then demonstrate
it via simulation in Section \ref{SimStudy}. The method is then applied
in Section \ref{Application} by estimating the excess infections, as
well as the duration of the excess, from 2020 election campaign rallies.
We end with a discussion of the results and our method in Section
\ref{Discussion}.

\section{Modeling and estimation methods}\label{Methods}

In this section, we start by giving relevant background information
about self-exciting point process models. We then describe our
adaptations to these models to estimate the excess number of points
induced by an event as well as the duration of time in which the excess
occurs.

\subsection{Self-exciting point processes}\label{sec:model}

Self-exciting point processes, also referred to as Hawkes processes
\citep{hawkes1971}, model the rate for points, \(x_i\)
\(i = 1, 2, \ldots, n\), accumulating in infinitesimally small regions
of time \[
\lambda(t|\mathcal{H}_t)  = \lim\limits_{\Delta t \to 0} \frac{E[N([t,t + \Delta t))|\mathcal{H}_t]}{\Delta t}
\] through the use of a counting measure, \(N(\cdot)\). The rate at
which points occur are typically modeled via their \emph{conditional
intensity} functions, \(\lambda(\cdot |\mathcal{H}_t)\), where
\(\mathcal{H}_t\) denotes the the historical occurrence of points from
an initial time, \(t_0\), up to time \(t\), \(t > t_0\)
\citep{daley2004}. For applications of points occurring in time, these
conditional intensity functions often take on the form \[
\lambda(t|\mathcal{H}_t) = \mu + \sum_{i:t_i < t} \nu(t - t_i)
\] where \(\mu\), here taken to be a positive-valued constant, describes
the stochastic occurrence of \emph{background points} (i.e.~parent
points not directly caused by previous occurrences), and
\(\nu(t - t_i)\) describes the triggering, or self-exciting, mechanism
of the process. The triggering function, \(\nu(t-t_i)\), is the
component responsible for the \emph{self-exciting} moniker. The function
describes how child points are produced, either from background points
or subsequent children of the background points. The triggering function
further defines how the self-excitation decays over time such that the
contagion-like process of a point does not continue indefinitely.

\subsection{Model adaptations for estimating effects of outside events}\label{sec:adaptations}

For this article, we utilize a conditional intensity function of the
form \begin{equation}\label{eq:ci}
\lambda_\ell(t|\mathcal{H}_{\ell,t}) = \mu_\ell + K_\ell(t) \sum_{i:t_i < t} g_\ell(t - t_i).
\end{equation} Here, \(\lambda_\ell(t|\mathcal{H}_{\ell,t})\) describes
the conditional intensity for the occurrence of points at time \(t\) and
location \(\ell\), \(\ell = 1, 2, \ldots, m\), given the history of
occurrences for location \(\ell\) up to time \(t\),
\(\mathcal{H}_{\ell,t}\). In the application we describe in Section
\ref{Application}, we model COVID-19 infections for the \(\ell^{th}\)
county in which a campaign rally was held for then President, Donald
Trump. Thus, \(m\) would denote the total number of counties in the
United States in which campaign rallies took place. We take \(\mu_\ell\)
to be the baseline rate of new points occurring in the \(\ell^{th}\)
location and \(K_\ell(t)\) to be the expected number of subsequent
points caused by previous ones, which we refer to as
\emph{productivity}. We let \(K_\ell(t)\) be a function of time, with
additional details given below. The triggering function,
\(g_\ell(t - t_i)\), describes how the rate of the occurrence of
additional points, caused by previous points, decays over time for
location \(\ell\). In this formulation of the conditional intensity
function, the parameter \(\mu_\ell\), the parameters of the function
\(K_\ell(t)\), and the function \(g_\ell(t-t_i)\) are considered unknown
and must be estimated from the data for location \(\ell\). Rather than
adopt a parametric form, such as exponential or power law decay, to
describe how the influence of previous points decays over time, the
triggering function, \(g_\ell(t-t_i)\), is estimated nonparametrically
using histogram estimators which we describe in section
\ref{sec:estimation}.

Time varying productivities, such as \(K_\ell(t)\) used in Equation
\ref{eq:ci}, have been explored to some degree in peer-reviewed
literature \citep{schoenberg2019} and in pre-print form
\citep{schoenberg2020}. The use of time varying productivites with the
intent of estimating the additional productivity linked to an external
event, to our knowledge, is a novel application; consequently, we
describe our further adaptions to the model in Equation \ref{eq:ci}
here.

Let \(K_\ell(t)\) be defined as \[
K_\ell(t) =  k_\ell + k_\ell^* \cdot I_{(t_\ell^* < t < t_\ell')} + k_\ell' \cdot I_{(t_\ell' < t)}
\] where \(I_{(\cdot)}\) represents the indicator function. The
parameter \(k_\ell\) is taken to be the \emph{baseline} expected number
of additional points that are caused by previous ones in location
\(\ell\). The parameter \(k_\ell^*\) describes the additional expected
number of points, in addition to the baseline \(k_\ell\), which are
caused by some known outside event taking place at time time
\(t_\ell^*\) in location \(\ell\). This additional expected rate of
points occurring will last for some unknown duration which ends at time
\(t_\ell'\), \(t_\ell' > t_\ell^*\). Finally, \(k_\ell'\) describes the
change in the base rate of expected points once the duration of the
extra productivity induced by the outside event has diminished. With
these modifications in place, we thus adopt the conditional intensity
function for location \(\ell\) as \begin{equation}\label{eq:ci1}
\lambda_\ell(t|\mathcal{H}_{\ell,t}) = \mu_\ell + (k_\ell + k_\ell^* \cdot I_{(t_\ell^* < t < t_\ell')} + k_\ell' \cdot I_{(t_\ell' < t)})  \sum_{i:t_i < t} g_\ell(t - t_i) 
\end{equation}

Using the conditional intensity function as proposed in Equation
\ref{eq:ci1}, we estimate constant values for the baseline productivity
of previous points, \(k_\ell\), the additional productivity induced by
some outside event occurring at some known time, \(k_\ell^*\), and how
the baseline rate of productivity has changed once the added
productivity associated with the outside event has waned, \(k_\ell'\).
We further estimate the end time of the additional productivity,
\(t_\ell'\).

By estimating \(k_\ell^*\) and \(t_\ell'\), we compute the expected
number of points which occurred in excess of the baseline as well as the
duration of time, \((t_\ell' - t_\ell^*)\), in which these additional
points occurred. This formulation is beneficial as, in instances where
the outside event does not add additional productivity over the baseline
rate, we would expect the model to produce estimates for \(k_\ell^*\)
and \((t_\ell' - t_\ell^*)\) which are close to zero. This expectation
then implies that we have a means to identify situations where the
outside event did not seem to have an identifiable impact on the rate of
points occurring for individual locations.

\subsection{Model estimation}\label{sec:estimation}

To estimate the model parameters and triggering function of Equation
\ref{eq:ci1}, we adopt, with several modifications described below, the
nonparametric Model Independent Stochastic Declustering (MISD) algorithm
proposed by Marsan and Lengliné \citep{marsan2008}. The MISD algorithm,
a variant of the EM-algorithm \citep{dempster1977}, allows for a causal
structure for the occurrence of points to be calculated
probabilistically using an iterative process to estimate the probability
that a point was caused by a previous point, or conversely is a
background point. These probabilities are then used to estimate the
constant values of step-functions for the triggering function, referred
to as histogram estimators, as well as parameters for the background
rate and productivity.

For the histogram estimators, the pairwise differences in times of the
events are computed, and then each inter-event time difference is placed
into a set of disjoint intervals or bins. Based on the probabilities of
the iterative process, constant values are estimated for each interval
in order to fit the model. The algorithm utilizes a lower-triangular
probability matrix, \(\mathbf{P}\), such that \(p_{ij}\) represents the
probability that event \(i\) is triggered by event \(j\), for \(i > j\),
while \(p_{ii}\) represents the probability that event \(i\) is a
background point. The algorithm iterates over a sequence of steps in
which estimates of the background rate, triggering functions, and
probability matrix are updated until convergence has been attained. A
more detailed description of the algorithm and standard error estimates
may be seen in \citet{Fox2016}.

In our approach, \(K_\ell(t)\) is a step-function with constants
\(\{k_\ell, k_\ell^*, k_\ell'\}\) estimated for time-periods
\(\{(t_{0_\ell}, t_\ell^*], (t_\ell^*, t_\ell'], (t_\ell', T_\ell]\}\),
respectively. Here, \(t_{0_\ell}\) denotes the starting point in which
we estimate the \emph{baseline} productivity for location \(\ell\),
\(t_\ell^*\) denotes when the outside event occurs, \(t_\ell'\) denotes
when the extra productivity ends that is induced by the event, and
\(T_\ell\) denotes the end of the time period. Each value of
\(K_\ell(t)\) is estimated using the parentage probability matrix with
\[
k_\ell = \sum_{j:t_j\in(t_a,t_b]} \sum_{i = j+1}^{N} \frac{p_{ij}}{(t_b - t_a)}
\] where \(t_a\) and \(t_b\) denote the beginning and ending times for
the time-periods
\(\{(t_{0_\ell}, t_\ell^*], (t_\ell^*, t_\ell'], (t_\ell', T_\ell]\}\).

While the starting time for the extra productivity induced by an event,
\(t_\ell^*\) is known, the end time of the extra productivity,
\(t_\ell'\), is not and must be estimated. To estimate \(t_\ell'\), we
start by setting \(t_\ell' = t_\ell^* + 1\), which can be thought of as
the extra productivity lasting only for a single day, and then iterate
through each possible length for the duration until
\(t_\ell' = T_\ell\), computing the value of \(k_\ell^*\) for each
iteration. We then fit a LOESS curve \citet{cleveland} to the durations
of the extra productivity and the estimates of \(k_\ell^*\) to determine
the day in which the first maximum occurs. Setting \(\hat{t}_\ell'\) to
be the date of the first maximum can be thought of as the date in which
the extra productivity begins to wane and the process begins to fall
back down to the baseline level of productivity.

To reduce the chance of our procedure erroneously specifying
\(\hat{t}_\ell'\) that is different from \(t_{\ell}^*\) when the event
did not in fact induce any extra productivity, we compare the average
number of points occurring just before the event to the average number
of points occurring just after the event using a two-sample t-test. In
our application, for example, we compared the COVID-19 cases for each
county in which then President Trump held a campaign rally and compared
the average number of cases which occurred two-weeks before the rally to
the average number of cases which occurred two-weeks after. In this
example, two-weeks before/after was selected roughly based on the time
in which an infected individual remains contagious \citep{cevik2021}. If
the t-test returns a non-significant p-value, we then state that the
event occurring at time \(t_\ell^*\) did not induce any extra
productivity.

\section{Simulation Study}\label{SimStudy}

In order to assess the performance of our modified model and MISD
algorithm, we implement a small simulation study. In this study, we are
specifically interested in examining how well our proposed method
detects the true duration of the increased productivity period as well
as its ability to estimate the average increase.

For our study, we simulate points for 100 days in 250 locations using
the method for simulating self-exciting point processes described in
\citet{zhuang2004}. We use a background rate of \(\mu = 3\) and an
exponential temporal decay with rate 0.25 to mimic the rate of decay we
roughly observe in our application (Section \ref{Application}).

For each location, an ``event'' will take place on the 31st day which
will increase the productivity of future points occurring for a duration
of 0, 10, 20, 30, or 40 days. After the duration of the extra
productivity ends, the productivity returns to a slightly higher than
baseline value, which again mimics what we found in our real-data
application. For simulated locations in which the duration of the
additional productivity is zero, we maintain the baseline level of
productivity which enables us to assess our method's ability to avoid
detecting erroneous added productivity when none is present.

The simulation method we employ simulates points in time as positive
real numbers, \(t_i \in (0, 100]\) and is parameterized using the
average number of child-points each individual point is expected to
produce. We aggregate the points occurring in each day to mimic the
case-counts data used in our application. In the simulations where the
event occurring on day 31 causes a temporary elevation in the occurrence
of points, we start with a baseline expected number of offspring points
of \(0.2\) which, once the event occurs, increases to \(1.0\), a
five-fold increase, before falling back down to \(0.4\) once the effect
of the event wanes, a two-fold increase relative to the baseline. For
simulations in which no productivity increase occurs, we set the
expected number of offspring to be \(0.2\) for each time period, before
and after the event. A summary of the scenarios used in our simulation
can be seen in Table \ref{tab:sim-summary}.

\begin{table}[h]

\caption{\label{tab:sim-summary}Summary of simulation scenario parameters}
\centering
\begin{tabular}[t]{ccc}
\toprule
Number of locations & Duration of extra productivity & Additional productivity\\
\midrule
50 & 0 & No increase\\
50 & 10 & 5-fold increase\\
50 & 20 & 5-fold increase\\
50 & 30 & 5-fold increase\\
50 & 40 & 5-fold increase\\
\bottomrule
\end{tabular}
\end{table}

The results from the simulation study pertaining to the estimation of
the duration for increased productivity are outlined in Table
\ref{tab:sim_duration}. For each duration, the minimum, \(5^{th}\)
percentile, \(25^{th}\) percentile, median, \(75^{th}\) percentile,
\(95^{th}\) percentile, and maximum error in the duration estimates are
computed by comparing the estimated duration of the extra productivity
to the actual duration,
\(\left(\hat{t}_\ell' - \hat{t}_\ell^*\right) - \left(t_\ell' - t_\ell^*\right)\).
Errors which are positive thus correspond to the method over-estimating
the duration and vice versa when the estimated duration under-estimates
the duration.

\begin{table}[h]

\caption{\label{tab:sim_duration}Differences between estimated and actual durations from simulations}
\centering
\begin{tabular}[t]{cccccccc}
\toprule
True Duration & Minimum & 5th & 25th & Median & 75th & 95th & Maximum\\
\midrule
0 & 0 & 0 & 0 & 0.0 & 0 & 9 & 26\\
10 & -10 & -5 & -1 & 1.0 & 2 & 3 & 5\\
20 & -9 & -5 & -1 & 1.0 & 2 & 3 & 4\\
30 & -7 & -5 & 0 & 1.0 & 2 & 3 & 4\\
40 & -4 & -2 & 0 & 1.5 & 3 & 3 & 3\\
\bottomrule
\end{tabular}
\end{table}

For instances in which the duration of additional productivity was 0
days, our method incorrectly estimated a duration greater than 0 four
times. When the duration was 10 days, we fail to detect the additional
productivity once. In the simulation scenarios with non-zero durations,
the median error was at most 1.5, with an overwhelming majority of
model-produced estimates within a few days of the true durations,
yielding an interquartile range of -5 days to +3 days at its largest.
Even at the extremes, the model estimates the duration to be within 10
days of the true value which is not entirely unexpected given the
variability of self-exciting point processes.

Table \ref{tab:sim_prod} shows the estimated additional productivity,
\(k_\ell^*\), for each set of 50 simulated locations for each duration
in which additional productivity was present. We again present the
minimum, maximum, and \(5^{th}\), \(25^{th}\), \(50^{th}\), \(75^{th}\),
\(95^{th}\) percentiles of the estimates. Note as well that, ideally,
the estimated additional productivity should be close to five as we
employed a five-fold increase in our simulations.

\begin{table}[h]

\caption{\label{tab:sim_prod}Estimated additional productivity relative to baseline from simulations}
\centering
\begin{tabular}[t]{cccccccc}
\toprule
Durations & Minimum & 5th & 25th & Median & 75th & 95th & Maximum\\
\midrule
10 & 0.00 & 0.83 & 1.27 & 1.77 & 2.37 & 3.70 & 23.01\\
20 & 0.00 & 1.35 & 2.52 & 3.27 & 4.47 & 5.44 & 18.30\\
30 & 0.94 & 1.28 & 3.08 & 4.26 & 5.21 & 6.22 & 6.97\\
40 & 0.85 & 1.59 & 3.71 & 5.73 & 6.58 & 9.02 & 10.54\\
\bottomrule
\end{tabular}
\end{table}

From the table, we see that the median estimated additional productivity
starts smaller than the five-fold increase we expected, for the 10-day
duration simulations, and gradually increases as the duration increases.
For the 40-day duration catalogs, the productivity increase is very near
what is expected with the middle 50\% of the estimates ranging from 3.71
to 6.58. Given that it takes some amount of time for the additional
productivity to fully develop in the simulations, we do not view the
estimates for the 10-day and 20-day durations as inaccurate. Rather, we
view them as being plausible estimates for the true amount of additional
productivity that could be realized over relatively short time-spans.
Additionally, we note that as the duration of the extra-productivity
time-period increases, the point process has a larger time span in which
to generate points, making the longer time-durations better at
estimating the five-fold increases used when generating the data. We
also notice that, as the duration increases, the 95th percentile also
grows which, again, is not unexpected given the inherent variability of
self-exciting point processes.

Overall, we find the ability of our proposed method to estimate the
duration and additional productivity of the simulated catalogs to be
rather strong, though we note that it is also capable of producing
inaccurate estimates on occasion. It is entirely plausible that altering
certain parameters of the modified MISD algorithm, such as changing the
LOESS smoothing bandwidth or the time intervals in which we estimate the
decay of the points, may have improved the estimates further. However,
we avoid doing this so as not to overfit the simulated catalogs of
points.

\section{Application: Estimating excess infections and their durations for 2020 election campaign rallies}\label{Application}

In this section, we apply our modified self-exciting point process
model, described in Section \ref{Methods}, to daily COVID-19 case counts
data for each county in which then President Donald Trump held a
campaign rally during the 2020 United States election cycle. We start
with a description of the data used in our analysis in Section
\ref{sec:data} and then describe our evaluation methods and results in
Section \ref{sec:evalmethods} and Section \ref{sec:appresults},
respectively.

\subsection{Data}\label{sec:data}

COVID-19 data for this project was taken from the \emph{New York Times}
[2021], reporting daily case counts in the United States at the
county level. Recorded COVID-19 cases in the United States began on
January 21, 2020 in Snohomish County, Washington, with subsequent case
counts being maintained daily. As county level reporting is the finest
resolution available, we use this data to track COVID-19 cases within
the counties, rather than cities, in which then President Trump held a
campaign rally during his 2020 re-election campaign. For each county,
data was drawn 30 days prior to the campaign rally in order to estimate
the baseline productivity of viral transmission. We then used data, for
each county, occurring 150 days after the rally was held to provide
sufficient time to encapsulate the duration of any additional
productivity induced by the campaign rally.

\begin{figure}
\centering
\includegraphics{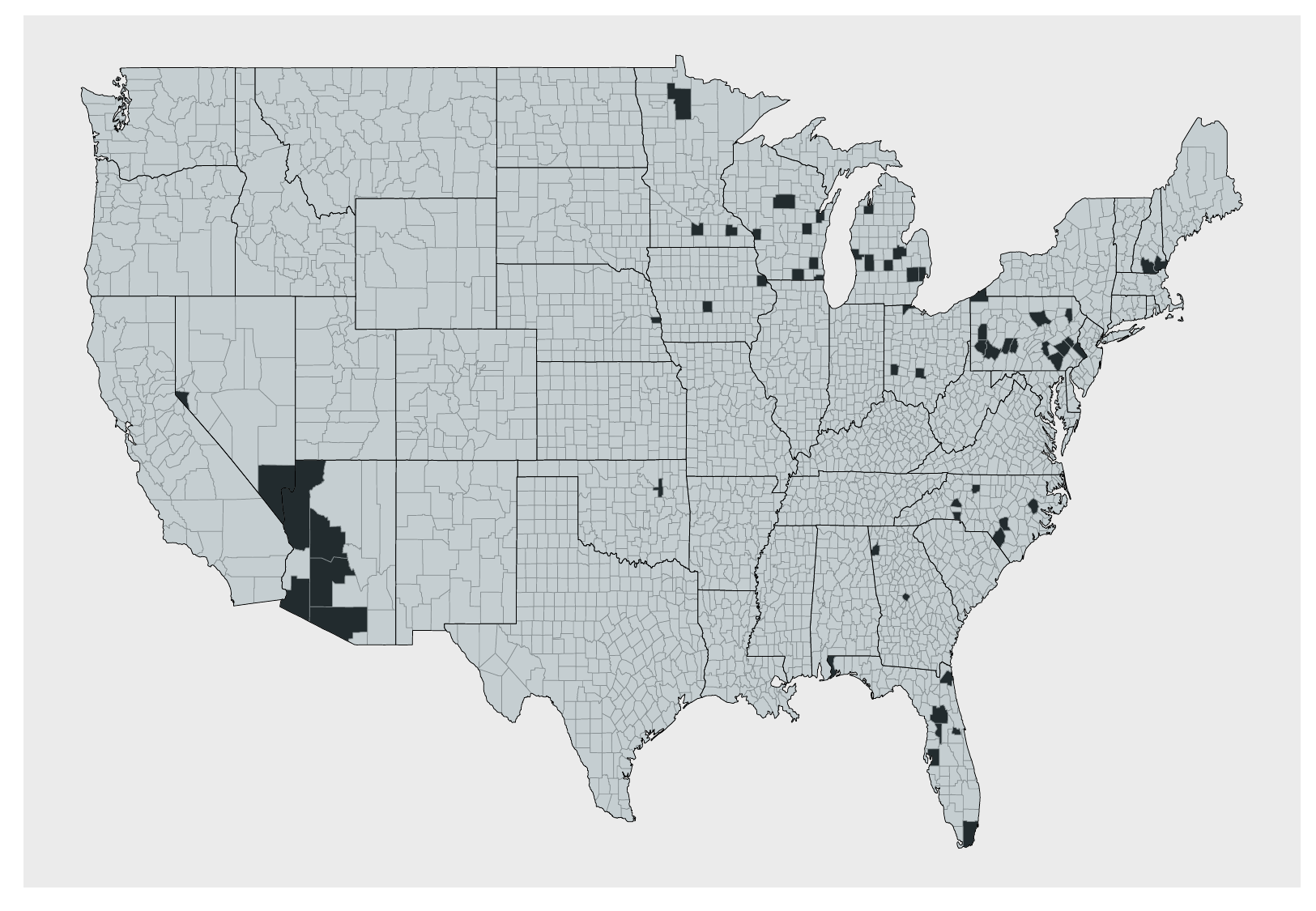}
\caption{\label{fig:county_map}Counties within the United States that
hosted a President Trump campaign rally in 2020.}
\end{figure}

In this application, we study the effects of rallies held during the
COVID-19 pandemic, beginning on June 20, 2020 and concluding on November
2, 2020. In total, 67 rallies were held on or after June 20; of these 67
rallies, 65 rallies were held in cities/towns that fall within
designated counties. President Trump held repeat rallies in three
counties (Cumberland, North Carolina; Lackawanna, Pennsylvania;
Maricopa, Arizona). For these three counties, only the first rally held
in each location will be considered as the during-rally effect of the
first rally may still be present at the time of the second rally.
Focusing on only the initial rally will also prevent the double counting
of potentially rally-related cases. After removing repeated rally
counties and non-county affiliated rallies, the remaining 62 rallies
will be used in this study and are highlighted in Figure
\ref{fig:county_map}.

Though official attendance numbers are not publicly available for any
rally, rough estimates have been reported by local news organizations.
Such estimates were obtained for each rally in order to verify that
attendance was large enough to potentially spur additional cases within
the county. A few smaller rallies hosted hundreds of attendees, while
most rallies were attended by thousands of individuals. The
self-exciting point process models introduced in Section \ref{Methods}
and further specified in Section \ref{sec:evalmethods} below do not
require attendance data, so the lack of precise records will not impede
analyses.

\subsection{Evaluation methods}\label{sec:evalmethods}

To verify whether an increase in cases at the county level may be
attributed to President Trump's campaign rallies, or was simply part of
a broader trend within the state, we will compare the increase of cases
seen within a rally county to the increase in cases within the state as
a whole. After applying the MISD algorithm to the county in which a
campaign rally took place, we then use the same modeling approach to the
respective state while specifying the same the pre-, during-, and
post-rally time-periods estimated by the county level model. To focus
more directly on the pattern of COVID cases happening in the state and
limit interaction with rallies, the statewide model includes case data
from only counties in which President Trump did not hold a rally at any
point.

Once the county and state models have been fit, we estimate the number
of cases that may be linked to President Trump's campaign rallies in the
following manner. We first find the population adjusted increase in
productivity at both the county and state levels for location \(\ell\)
and produce the multiplier, \(\kappa_\ell\), such that

\begin{equation*}
\kappa_{\ell} = \frac{k_{\ell_\text{county}}^* - k_{\ell_{\text{county}}}}{\text{population}_{\ell_\text{county}}} - \frac{k_{\ell_\text{state}}^* - k_{\ell_\text{state}}}{\text{population}_{\ell_\text{state}}},
\end{equation*}

\noindent where the subscripts ``county'' and ``state'' denote county
and state level estimates for location \(\ell\), respectively. We may
then find the estimated number of cases resulting from a rally, above
what was observed in the rest of the state, via
\begin{equation}\label{eq:kappa}
\text{cases}_\ell = \kappa_\ell \times \text{population}_{\ell_{\text{county}}} \times \left(t^{'}_{\ell} - t^{*}_{\ell}\right)
\end{equation} where \((t^{'}_{\ell} - t^*_\ell)\) defines the duration
(in days) in which the rally county experiences a heightened
productivity.

\subsection{Application results}\label{sec:appresults}

Table \ref{tab:results_success} displays the 34 campaign rallies that
were found to have a significant increase of COVID-19 cases following a
campaign rally, which we refer to as a \emph{rally effect}, relative to
the rest of the state. The table includes the county and state in which
the rally was held and the date the rally took place. It reports the
estimated background rate of infection (\(\mu_\ell\)), the baseline
productivity of additional infections before the rally took place
(\(k_\ell\)), the estimated additional productivity of infections linked
to the campaign rally (\(k_\ell^*\)), and the estimated number of days
the rally effect persisted as well as the estimated number of cases
resulting from the rally based on Equation \ref{eq:kappa}.

\begin{table}[h]

\caption{\label{tab:results_success}Counties in which an increase in COVID-19 productivity, over baseline, was found following campaign rallies held by then President Donald Trump during the 2020 United States election cycle}
\centering
\begin{tabular}[t]{lllrrrrr}
\toprule
County & State & Date of Rally & $\mu_\ell$ & $k_\ell$ & $k_\ell^*$ & Duration & Cases\\
\midrule
Tulsa & OK & 2020-06-20 & 57.43 & 0.00 & 183.40 & 50 & 4274\\
Blue Earth & MN & 2020-08-17 & 7.01 & 6.25 & 17.07 & 22 & 315\\
Marathon & WI & 2020-09-17 & 6.69 & 0.12 & 159.75 & 61 & 8602\\
Beltrami & MN & 2020-09-18 & 1.53 & 0.79 & 29.40 & 87 & 1362\\
St. Louis & MN & 2020-09-30 & 4.81 & 35.20 & 148.57 & 59 & 4334\\
\addlinespace
Cambria & PA & 2020-10-13 & 4.82 & 1.04 & 124.57 & 64 & 5353\\
Polk & IA & 2020-10-14 & 34.09 & 58.81 & 431.27 & 40 & 10676\\
Muskegon & MI & 2020-10-17 & 4.66 & 6.86 & 190.78 & 31 & 4690\\
Rock & WI & 2020-10-17 & 6.99 & 51.27 & 174.50 & 29 & 2743\\
Yavapai & AZ & 2020-10-19 & 7.88 & 5.81 & 148.44 & 75 & 3603\\
\addlinespace
Pima & AZ & 2020-10-19 & 38.44 & 24.15 & 818.36 & 95 & 25203\\
Erie & PA & 2020-10-20 & 10.69 & 0.09 & 154.64 & 69 & 3334\\
Escambia & FL & 2020-10-23 & 10.26 & 16.70 & 269.84 & 96 & 4023\\
Pickaway & OH & 2020-10-24 & 2.89 & 13.09 & 50.20 & 79 & 834\\
Waukesha & WI & 2020-10-24 & 11.92 & 161.38 & 883.63 & 30 & 18765\\
\addlinespace
Hillsborough & NH & 2020-10-25 & 24.82 & 0.49 & 254.46 & 88 & 5159\\
Lehigh & PA & 2020-10-26 & 23.22 & 0.17 & 262.07 & 81 & 7509\\
Lancaster & PA & 2020-10-26 & 37.49 & 12.67 & 291.96 & 65 & 3905\\
Blair & PA & 2020-10-26 & 3.57 & 14.19 & 122.92 & 53 & 3527\\
Clinton & MI & 2020-10-27 & 10.66 & 3.79 & 101.86 & 20 & 1627\\
\addlinespace
La Crosse & WI & 2020-10-27 & 5.59 & 41.96 & 146.63 & 16 & 1346\\
Douglas & NE & 2020-10-27 & 25.29 & 203.08 & 787.79 & 25 & 11056\\
Mohave & AZ & 2020-10-28 & 6.75 & 0.06 & 182.05 & 81 & 6510\\
Oakland & MI & 2020-10-30 & 124.66 & 73.24 & 1257.86 & 17 & 15342\\
Olmsted & MN & 2020-10-30 & 7.55 & 23.08 & 169.99 & 22 & 2313\\
\addlinespace
Bucks & PA & 2020-10-31 & 47.23 & 0.19 & 355.42 & 52 & 6205\\
Berks & PA & 2020-10-31 & 50.76 & 37.22 & 274.95 & 77 & 3242\\
Butler & PA & 2020-10-31 & 15.57 & 5.36 & 147.12 & 57 & 3773\\
Lycoming & PA & 2020-10-31 & 6.57 & 0.83 & 94.47 & 59 & 2872\\
Macomb & MI & 2020-11-01 & 141.74 & 75.83 & 1059.51 & 13 & 10301\\
\addlinespace
Catawba & NC & 2020-11-01 & 7.35 & 45.56 & 132.87 & 82 & 1052\\
Miami-Dade & FL & 2020-11-01 & 351.71 & 165.67 & 2312.79 & 75 & 67006\\
Grand Traverse & MI & 2020-11-02 & 6.93 & 10.29 & 57.94 & 29 & 570\\
Kenosha & WI & 2020-11-02 & 5.03 & 74.82 & 204.56 & 20 & 1629\\
\bottomrule
\end{tabular}
\end{table}

Table \ref{tab:success_summ} shows the minimum, maximum, and the
\(5^{th}\), \(25^{th}\), \(50^{th}\), \(75^{th}\), \(95^{th}\)
percentiles of the estimated additional productivity relative to the
baseline, the duration of the effect, and the additional number of cases
relative to the rest of the state for the counties in which contained a
rally-effect.

\begin{table}[h]

\caption{\label{tab:success_summ}Estimated additional productivity relative to baseline from simulations}
\centering
\begin{tabular}[t]{>{\raggedright\arraybackslash}p{1.25in}ccccccc}
\toprule
 & Minimum & 5th & 25th & Median & 75th & 95th & Maximum\\
\midrule
Additional productivity relative to baseline & 1.29 & 2.54 & 3.39 & 6.09 & 11.11 & 22.03 & 26.73\\
Duration & 13.00 & 16.65 & 29.00 & 58.00 & 76.50 & 90.45 & 96.00\\
Additional cases & 315.00 & 741.60 & 2420.50 & 3964.00 & 7259.25 & 21018.30 & 67006.00\\
\bottomrule
\end{tabular}
\end{table}

For counties in which a rally effect was detected, there was a 6-fold
increase in the median productivity of COVID-19 cases with an
inter-quartile range of 7.72. The median duration of this increase was
roughly two months with a minimum duration of about two weeks and a
maximum increase of about three months. Additional cases, which we
attribute to the rally after taking into account the changes seen at the
state level, ranged from a few hundred up to tens of thousands with a
median estimate of about 4000 additional cases per rally.

The 28 counties in which campaign rallies were estimated to have little
to no impact by our model, or were estimated to have little to no impact
relative to the rest of the state, are shown in Table
\ref{tab:results_fail}. The table also includes counties for which the
model estimation processes failed to converge, which is a known issue
for self-exciting processes \citep{veen2008}. The information contained
in the table is similar to that of Table \ref{tab:results_success}
except for the additional column which describes whether the model
estimation process failed to converge.

\begin{table}[h]

\caption{\label{tab:results_fail}Counties in which little to no increase in COVID-19 productivity, over baseline, was found following campaign rallies held by then President Donald Trump during the 2020 United States election cycle}
\centering
\begin{tabular}[t]{llllllllc}
\toprule
County & State & Date of Rally & $\mu_\ell$ & $k_\ell$ & $k_\ell^*$ & Duration & Cases & Converged?\\
\midrule
Maricopa & AZ & 2020-06-23 & - & - & - & - & - & No\\
Winnebago & WI & 2020-08-17 & 15.23 & 0.03 & 0 & 0 & 0 & Yes\\
Yuma & AZ & 2020-08-18 & - & - & - & - & - & No\\
Lackawanna & PA & 2020-08-20 & 17.29 & 0 & 0 & 0 & 0 & Yes\\
Rockingham & NH & 2020-08-28 & 26.89 & 0 & 0 & 0 & 0 & Yes\\
\addlinespace
Westmoreland & PA & 2020-09-03 & 12.61 & 0 & 0 & 0 & 0 & Yes\\
Forsyth & NC & 2020-09-08 & 52.05 & 0.07 & 0 & 0 & 0 & Yes\\
Saginaw & MI & 2020-09-10 & 23.48 & 0.03 & 0 & 0 & 0 & Yes\\
Douglas & NV & 2020-09-12 & 3.06 & 0 & 0 & 0 & 0 & Yes\\
Clark & NV & 2020-09-13 & 252.04 & 131.97 & 0 & 0 & 0 & Yes\\
\addlinespace
Cumberland & NC & 2020-09-19 & 42.79 & 13.57 & 0 & 0 & 0 & Yes\\
Montgomery & OH & 2020-09-21 & 36.32 & 64.87 & 0 & 0 & 0 & Yes\\
Lucas & OH & 2020-09-21 & 34.61 & 0.57 & 0 & 0 & 0 & Yes\\
Allegheny & PA & 2020-09-22 & 65.1 & 0.16 & 0 & 0 & 0 & Yes\\
Duval & FL & 2020-09-24 & 99.78 & 49.46 & 0 & 0 & 0 & Yes\\
\addlinespace
Dauphin & PA & 2020-09-26 & 10.25 & 8.87 & 113.52 & 115 & 0 & Yes\\
Seminole & FL & 2020-10-12 & 30.51 & 0.07 & 147.76 & 106 & 0 & Yes\\
Pitt & NC & 2020-10-15 & 11.78 & 37.65 & 0 & 0 & 0 & Yes\\
Marion & FL & 2020-10-16 & 23.49 & 13.7 & 0 & 0 & 0 & Yes\\
Bibb & GA & 2020-10-16 & 9.44 & 13.99 & 0 & 0 & 0 & Yes\\
\addlinespace
Gaston & NC & 2020-10-21 & 15.74 & 64.38 & 0 & 0 & 0 & Yes\\
Sumter & FL & 2020-10-23 & 6.85 & 20.41 & 0 & 0 & 0 & Yes\\
Robeson & NC & 2020-10-24 & 6.63 & 49.83 & 0 & 0 & 0 & Yes\\
Hillsborough & FL & 2020-10-29 & 109.55 & 107.71 & 631.31 & 84 & 0 & Yes\\
Brown & WI & 2020-10-30 & 14.68 & 432.95 & 0 & 0 & 0 & Yes\\
\addlinespace
Dubuque & IA & 2020-11-01 & 3.85 & 121.99 & 0 & 0 & 0 & Yes\\
Floyd & GA & 2020-11-01 & 8.26 & 36.13 & 42.04 & 10 & 0 & Yes\\
Kent & MI & 2020-11-02 & - & - & - & - & - & No\\
\bottomrule
\end{tabular}
\end{table}

In Table \ref{tab:results_fail}, we see that there were three instances
when the estimation method failed to converge, specifically for Maricopa
County, Arizona, Yuma County, Arizona, and Kent County, Michigan. These
failures were due to some volatility in the case counts occurring in the
30 days prior to the campaign rally as we were able to get the models to
converge when using data from 32 to 35 days prior to the rally. To keep
our comparisons similar, we omit the estimates based on these lengthend
before-rally time periods. We further note that there were 22 instances
where the model detected no rally-effect and four instances in which the
estimated rally-effect was not distinguisable compared to the case
counts from the rest of the state containing the county.

\begin{figure}
\centering
\includegraphics{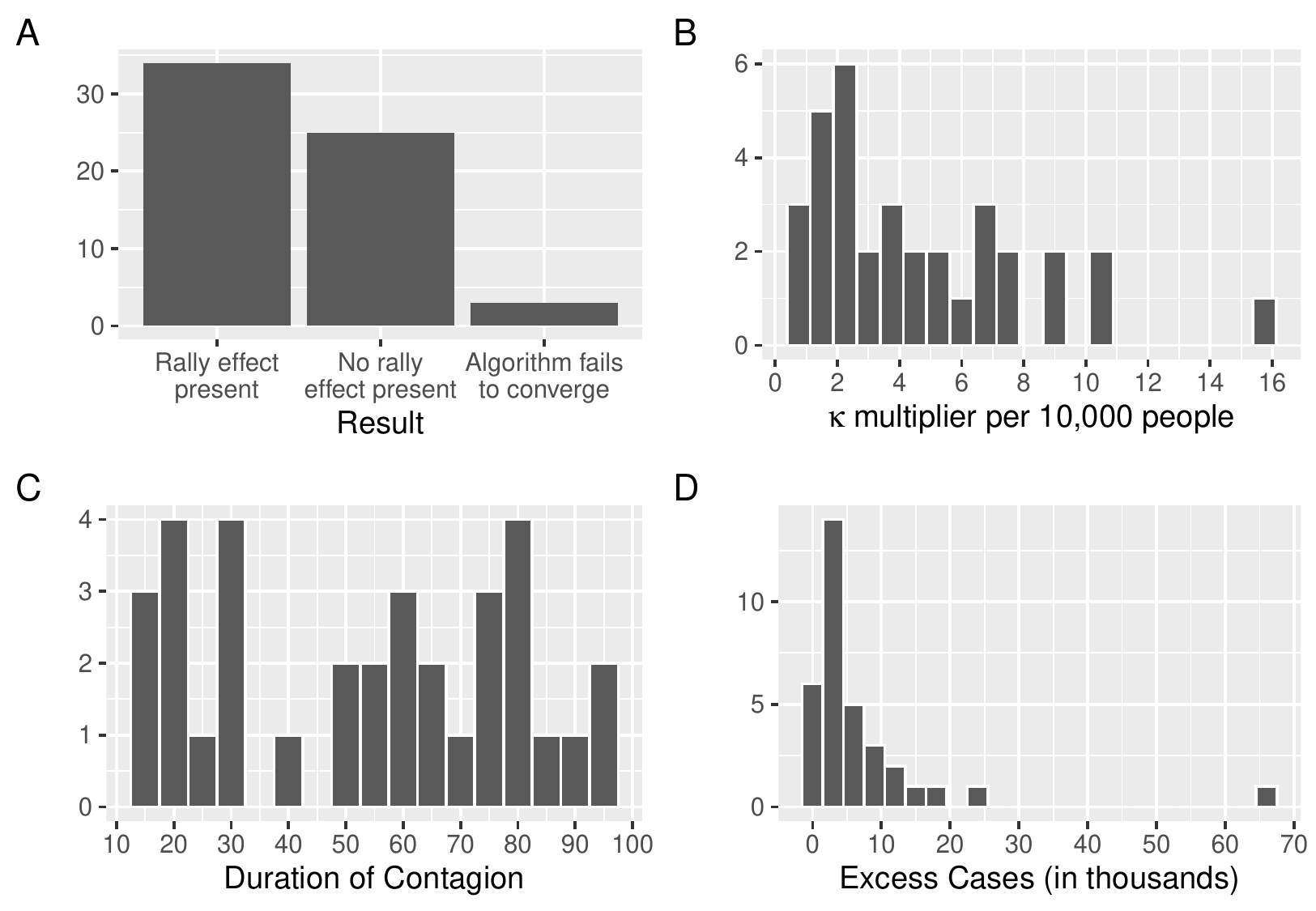}
\caption{\label{fig:plot_grid}(A) Distribution of county level results
in which the rally may have contributed to an increase in cases beyond
the state level, had no impact on cases, increased cases but at a rate
lower than state increases, or the algorithm failed to converge. (B)
Distribution of the multiplier seen in Equation 3. (C) Distribution of
the duration in which the additional contagion is estimated to last
following the rally. (D) Distribution of the number of excess cases that
may be attributed to President Trump's rallies.}
\end{figure}

\begin{figure}
\centering
\includegraphics{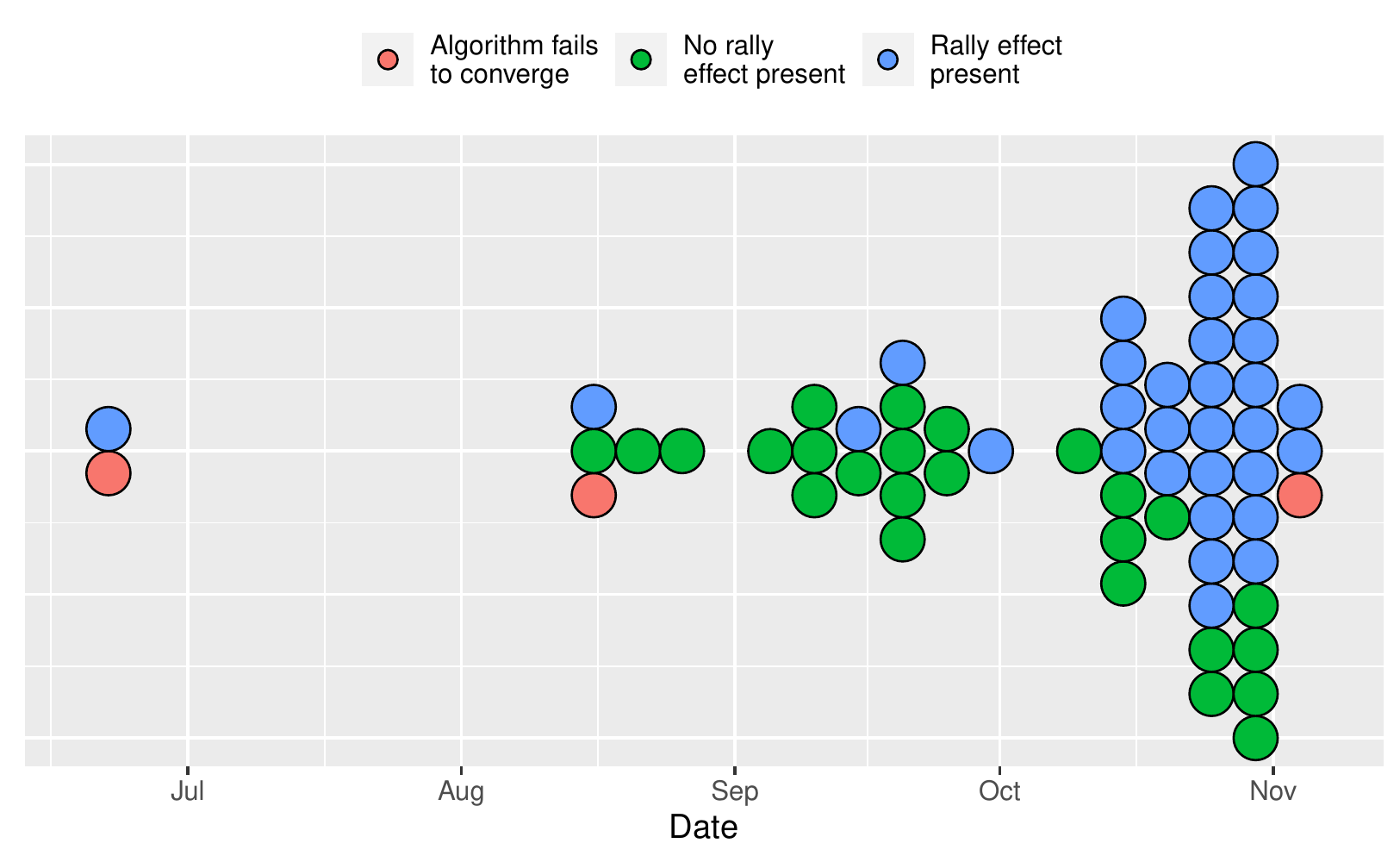}
\caption{\label{fig:plot_timeline}Model estimation outcomes by date of
rally. Please note that rallies are binned into 5-day intervals for
plotting purposes.}
\end{figure}

Figure \ref{fig:plot_grid} provides distributions of several results,
namely the number of cases, the \(\kappa\) multiplier seen in Equation
\ref{eq:kappa}, the duration of the excess contagion, and the overall
assessment of the rally on COVID-19 cases. Figure
\ref{fig:plot_timeline} then shows the results of the model estimation
over the time in which campaign rallies took place.

In Figure \ref{fig:plot_timeline}, we note that between August and
October, the majority of the campaign rallies were estimated to have no
impact on the daily COVID-19 case numbers or were estimated to be no
different than what was observed at the state level. Counties in which a
rally effect was found during these months tended to have populations
which were approximately five-times smaller, on average, than counties
where no effects were present (mean populations of 110,000 vs.~550,000,
respectively). We postulate that, given the state of the epidemic during
those months, the model was more adept at detecting additional
productivity in these smaller counties than in counties with larger
populations.

\section{Discussion}\label{Discussion}

In this article, we adapt temporal self-exciting point process models
and their nonparametric estimation techniques to estimate the duration
and effects that events, occurring at some time and place, have on the
productivity of the occurrence of subsequent points. In doing so, we
develop an alternative modeling approach to the synthetic control
methods which are commonly employed to estimate such effects. In our
approach, however, we avoid the necessity of finding suitable controls
and instead estimate the effects and duration using the data from the
location of interest directly.

We apply our methods to study the impact that campaign rallies held by
then President Donald Trump had on COVID-19 cases in counties across the
United States. By treating daily COVID-19 case counts as a self-exciting
point process, we are able to model the change in contagion
productivity, if any, which occurred after the rallies. And by fitting
models at both the county and state level, we may then estimate the
relative number of COVID-19 cases that may be linked to the rally.

In making comparisons between counties in which rallies were held and
the states containing the counties, we assume that similar state level
pandemic policies would apply uniformly throughout the state; however,
we recognize that regulations, political, and social adherence to such
policies vary widely across an individual state's counties
\citep{brzezinski2021}.

While our application demonstrates the method's ability to detect
increases in the occurrence of points, or COVID-19 cases, the approach
would be equally well-suited for applications in which a reduction in
the occurrence of points is thought to occur. Our methods are also
relatively simple compared to regression modeling techniques, needing
only the daily case counts to fit the model.

Similar to the findings of \citet{bernheim2020_real}, and in contrast to
the findings of \citet{Dhaval2021}, we find that the campaign rally
which took place in Tulsa, Oklahoma on June 20, 2020, did lead to an
increase in COVID-19 cases for that county. Further,
\citet{bernheim2020_real}, who used synthetic control methods to
estimate the campaign rally-related increase in COVID-19 infections for
18 rallies held between June 20, 2020 and September 30, 2020, estimated
the overall number of rally related cases to be approximately 30,000
additional infections. This leads to an estimated increase of
approximately 1600 additional cases per rally. Using our proposed
methods, we find that, on average in the counties for which an increase
in infection productivity was detected, there was approximately 750
additional cases per rally. Our method, however, compares the estimated
increase in case counts within the counties to their respective states,
rather than to the synthetic controls. We also estimate the duration of
the additional productivity rather than comparing infection rates for a
10-week period following the rally.

Overall, we estimate the total number of COVID-19 cases that may be
attributed to the 62 campaign rallies used in our study to be 253,055.
We further observe that 34 of 62 rallies were estimated to yield an
excess of COVID-19 cases while 25 rallies showed no evidence of increase
in case numbers.

\bibliographystyle{plainnat}
\bibliography{references.bib}

\end{document}